\documentclass[
    ,final            % use final for the camera ready runs
  ]
  {aipproc}

\layoutstyle{6x9}

\begin{document}

\title{Radio Recombination Lines from Starbursts: 
NGC 3256, NGC 4945 and the Circinus Galaxy}

\author{A. L. Roy}{
  address={Max-Planck-Institut f\"ur Radioastronomie, Auf dem
              H\"ugel 69, 53121 Bonn, Germany}
}

\author{W. M. Goss}{
  address={NRAO, PO Box O, Socorro, NM 87801, USA}
}

\author{Niruj R. Mohan}{
  address={Raman Research Institute, CV Raman Ave,
              Sadashivanagar, Bangalore 560080, India}
}

\author{T. Oosterloo}{
  address={ASTRON, PO Box 2, 7990 AA, Dwingeloo, The Netherlands}
}

\author{K.R. Anantharamaiah}{
  address={Raman Research Institute, CV Raman Ave,
              Sadashivanagar, Bangalore 560080, India; (deceased)}
}

\begin{abstract}
  
A renewed attempt to detect radio recombination lines from external galaxies
has resulted in the measurement of lines from several bright starburst
galaxies. The lines are produced by hydrogen ionized by young, high-mass
stars and are diagnostic of the conditions and gas dynamics in the starburst
regions without problems of dust obscuration. We present here detections of
the lines H91$\alpha$ and H92$\alpha$ near 8.6 GHz from the
starburst nuclei in NGC 3256, NGC 4945, and the Circinus galaxy 
using the ATCA and VLA.  Modelling the line emitting region
as a collection of H~II regions, we derive the required
number of H\,II regions, their temperature, density, and distribution.

\end{abstract}

\maketitle

%%%%%%%%%%%%%%%%%%%%%%%%%%%%%%%%%%%%%%%%%%%%
%% MAINMATTER
%%%%%%%%%%%%%%%%%%%%%%%%%%%%%%%%%%%%%%%%%%%%

\section{Introduction}

Radio recombination lines (RRLs) from nuclear starbursts pass unattenuated
through dust and so their strength tells about conditions in the starburst
without uncertainty caused by unknown
extinction that troubles optical and near-infrared studies.  

The potential for detecting extragalactic RRLs was first pointed out by
\cite{Shaver1978} and shortly thereafter RRLs were detected from the
starbursts in M\,82 and NGC\,253 (\cite{Shaver1977},
\cite{SeaquistBell1977}).  These two galaxies have since been studied at many
frequencies, yielding the physical state and kinematics in the nuclear regions
(e.g. \cite{AnantharamaiahGoss1997}, \cite{RodriguezRicoEtAl2004}).

Then came a period of surveys that yieled no further detections (
\cite{ChurchwellShaver1979}, \cite{BellSeaquist1978}, \cite{BellEtAl1984}).  A
renewed effort during the early 1990s using the Very Large Array (VLA) with
improved sensitivity detected RRLs near 8.6~GHz from several bright starburst
galaxies at a level $10\times$ weaker than the first two detections.  These
new detections are NGC 660, NGC 1365, NGC 2146, NGC 3628, NGC 3690, NGC~5253,
M 83, IC 694, Arp 220, Henize~2-10 (\cite{AnantharamaiahEtAl1993},
\cite{ZhaoEtAl1996}, \cite{ PhookunEtAl1998}, \cite{MohanEtAl2001}), NGC~1808
(\cite{Mohan2002}), and NGC~4945 at mm wavelengths
(\cite{ViallefondPrivComm}).

We present here three new detections: NGC 3256, NGC 4945 and the Circinus
galaxy.

\section{New ATCA and VLA Observations}

We selected five nearby (< 40 Mpc), infra-red luminous ($L_{\rm{FIR}} >
10^{10}~L_{\odot}$), dusty ($L_{\rm{FIR}} / L_{\rm{optical}} > 2$)
bright ($S_{\rm{100 \mu m}} > 10$~Jy) starburst galaxies (NGC 3256, NGC 4945,
NGC 6221, NGC 7552 and the Circinus Galaxy) to survey for RRL emission.
We also selected two nearby bright type 2 Seyfert galaxies (IC 5063 and
Fairall 49) to see whether the ionized gas in the narrow-line region
might also be detected.

We observed each galaxy for 10~h to 30~h with the ATCA \footnote{The Australia
Telescope Compact Array is part of the Australia Telescope, which is funded
by the Commonwealth of Australia for operation as a National Facility
managed by CSIRO.}  and made a confirmation observation of NGC~3256 with the
VLA \footnote{The National Radio Astronomy Observatory is a facility of the
National Science Foundation operated under cooperative agreement by
Associated Universities, Inc.}.  With the ATCA we observed simultaneously
the lines H91$\alpha$ and H92$\alpha$ near 8.6\,GHz.  With the VLA we observed
H92$\alpha$.

A bandpass calibrator was observed every few hours and phase corrections
obtained from self calibration of the continuum source were applied to the
spectral line data.  We subtracted the continuum emission using a linear fit
to each baseline spectrum using with UVLSF (\cite{CornwellUsonHaddad1992}).
The final images were made using natural or robust weighting to achieve
near-maximum possible signal-to-noise ratio.

We did not detect H91$\alpha$ + H92$\alpha$ emission
from NGC 6221, NGC 7552, IC 5063 and Fairall 49 after 10~h integrations
reaching rms of 0.35\,mJy\,beam per 1\,MHz channel.

\section{Detection of NGC 3256}

NGC 3256 is a pair of colliding disk galaxies 37\,Mpc distant (for 
$H_0 = 75$ km s$^{-1}$ Mpc$^{-1}$) that are merging and
display spectacular tidal tails (e.g. \cite{
Schweizer1986}).  The FIR luminosity is
$1.9\times10^{11} L_{\odot}$ following the method of \cite{HelouEtAl1985},
and the molecular gas mass, from CO emission, is large
($3\times10^{10} M_{\odot}$, \cite{SargentEtAl1989}).  The Br$\gamma$ and
[Fe~II] luminosities imply a total star formation rate of
$3.9~M_{\odot}~\mathrm{yr}^{-1}$.  VLA observations
at 6 cm with 4" resolution by \cite{SmithKassim1993} show emission over 30"
and arms of diffuse emission extending out towards the giant tidal arms seen
in H\,I (\cite{EnglishEtAl2003}).  At higher resolution (2"),
\cite{NorrisForbes1995} resolve the nucleus into two equal components; the
nuclei of the progenitor galaxies are both undergoing starbursts.

Our ATCA discovery observation is shown in Fig 1.  Integration time was
24.2\,h with 1\,MHz (35\,km\,s$^{-1}$) channels. The 8.4\,h VLA
follow-up observation with sensitivity of 0.081\,mJy\,beam$^{-1}$\,
channel$^{-1}$ yielded an essentially similar result and is
shown in \cite{RoyEtAl2004}.

\begin{figure}
\includegraphics[width=7cm]{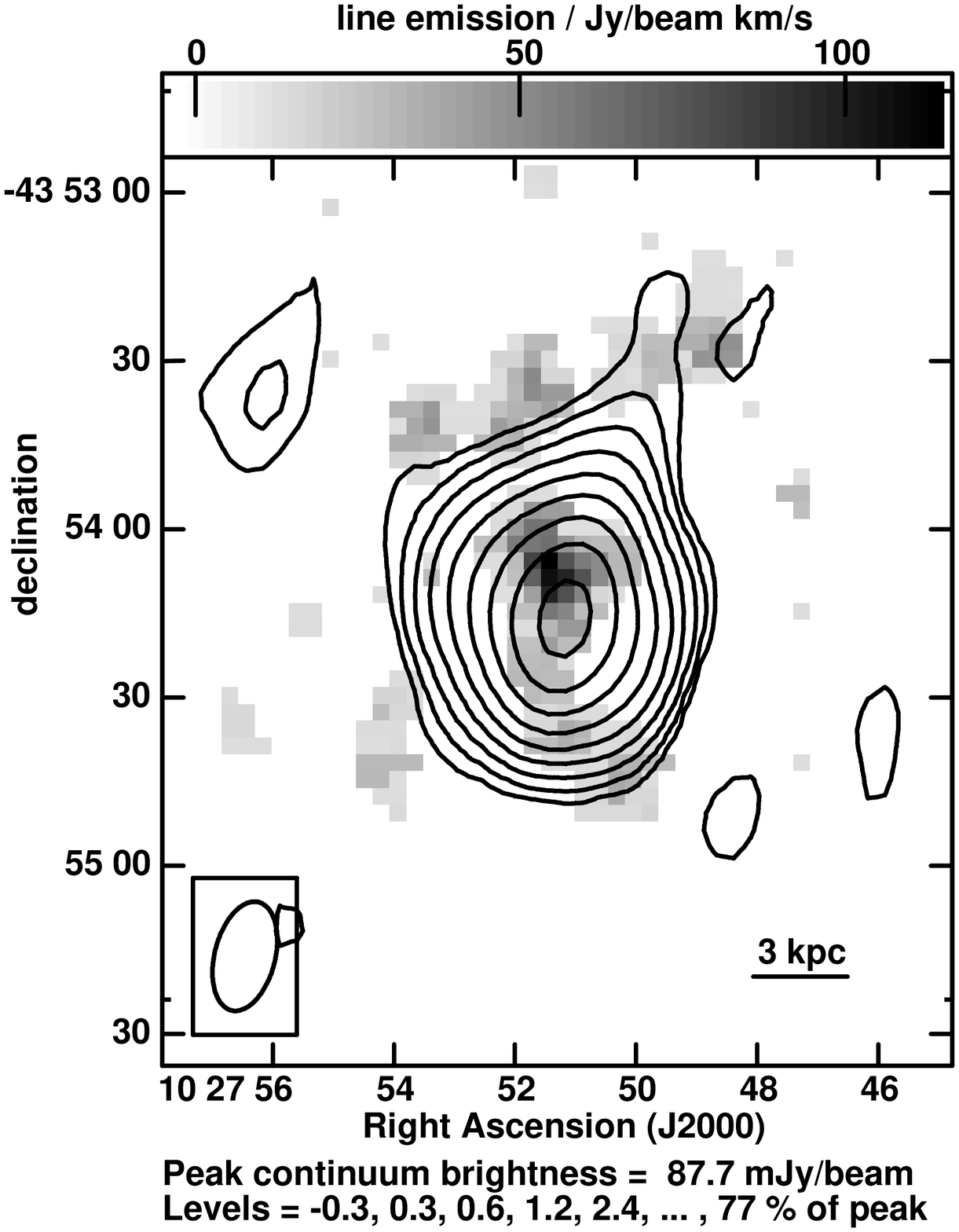}
\includegraphics[width=7cm]{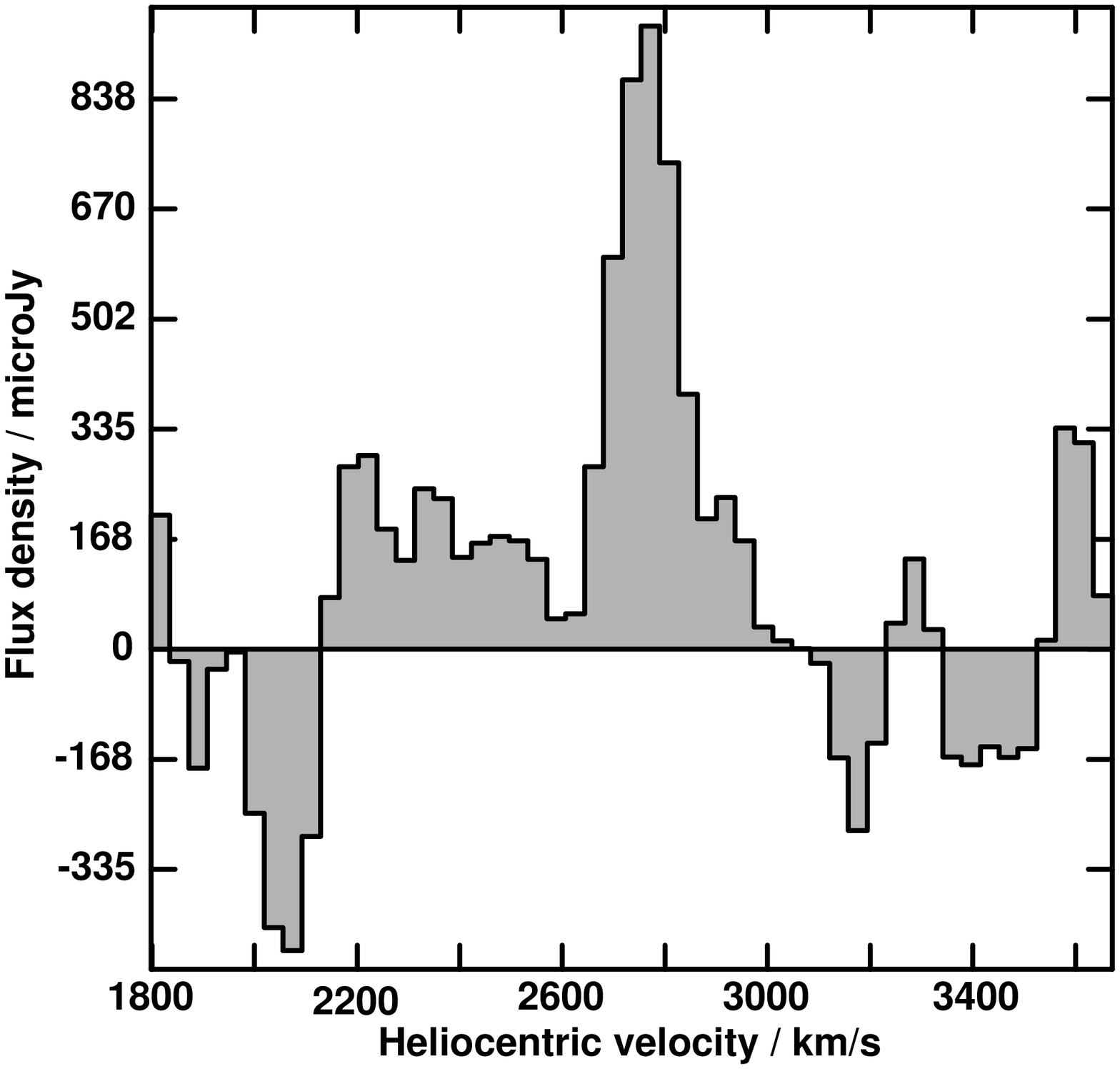}
\caption{Left:
ATCA 8.4 GHz continuum image of NGC 3256 observed 1994 Oct + 1995 Aug
(contours), superimposed on the grey scale zeroth-moment image showing
H91$\alpha$ + H92$\alpha$ line emission.  Beamsize is
$16.4'' \times 9.6''$. The offset between the line and continuum peaks
is perhaps significant as it was seen also in the VLA observation.
Right: ATCA H91$\alpha$ + H92$\alpha$ line profile integrated over the
line-emitting region in NGC 3256, observed 1994 Oct + 1995 Aug.
RMS noise is 0.14\,mJy\,beam$^{-1}$\,channel$^{-1}$.
}
\end{figure}

We used the observed line strength (1.0~mJy), line width (160~km~s$^{-1}$),
size of the line-emitting region (630\,pc), continuum emission (116\,mJy) and
spectral index (-0.7) to constrain conditions in the ionized gas.  We
considered a simple model consisting of a collection of spherical H~II
regions, all with the same diameter, electron temperature, $T_{\rm{e}}$,
electron density, $n_{\rm{e}}$, in a volume of uniform synchrotron
emission, following \cite{AnantharamaiahEtAl1993}.  Models
with 10 to 300 H\,II regions, all with $T_{\mathrm{e}} \sim 5000$~K,
$n_{\mathrm{e}} \sim (10^{3}$ to $10^{4})$\,cm$^{-3}$ and size of 
(5 to 600)\,pc produced good matches to the line and continuum emission.

The inferred mass of ionized gas is $4\times10^{4}~M_{\odot}$ to
$2\times10^{5}~M_{\odot}$, depending on the model conditions, which requires a
Lyman continuum flux of $2\times10^{52}$~s$^{-1}$ to $6\times10^{53}$~s$^{-1}$
to maintain the ionization.  This flux is equivalent to the Lyman continuum
output of 600 to 17000 stars of type O5, which infers a star-formation rate of
(0.5 to 14) $M_\odot$ yr$^{-1}$ when averaged over the lifetime of OB stars.

\section{Detection of NGC 4945}

NGC 4945 is a nearby (3.9 Mpc\cite{BergmanEtAl1992}) edge-on spiral galaxy
containing an obscured nuclear starburst and is the second-brightest Seyfert 2
galaxy in the X-ray sky at 100\,keV \cite{DoneEtAl1996}.  Its nucleus 
hosts water masers with up to 10\,Jy \cite{WhiteoakGardner1986}.

Our ATCA discovery observation is shown in Figs 2 and 3.  Integration time
was 22.7\,h with 35\,km\,s$^{-1}$\,channel$^{-1}$ and sensitivity of
0.16\,mJy\,beam$^{-1}$\,channel$^{-1}$.

\begin{figure}
\includegraphics[width=5cm]{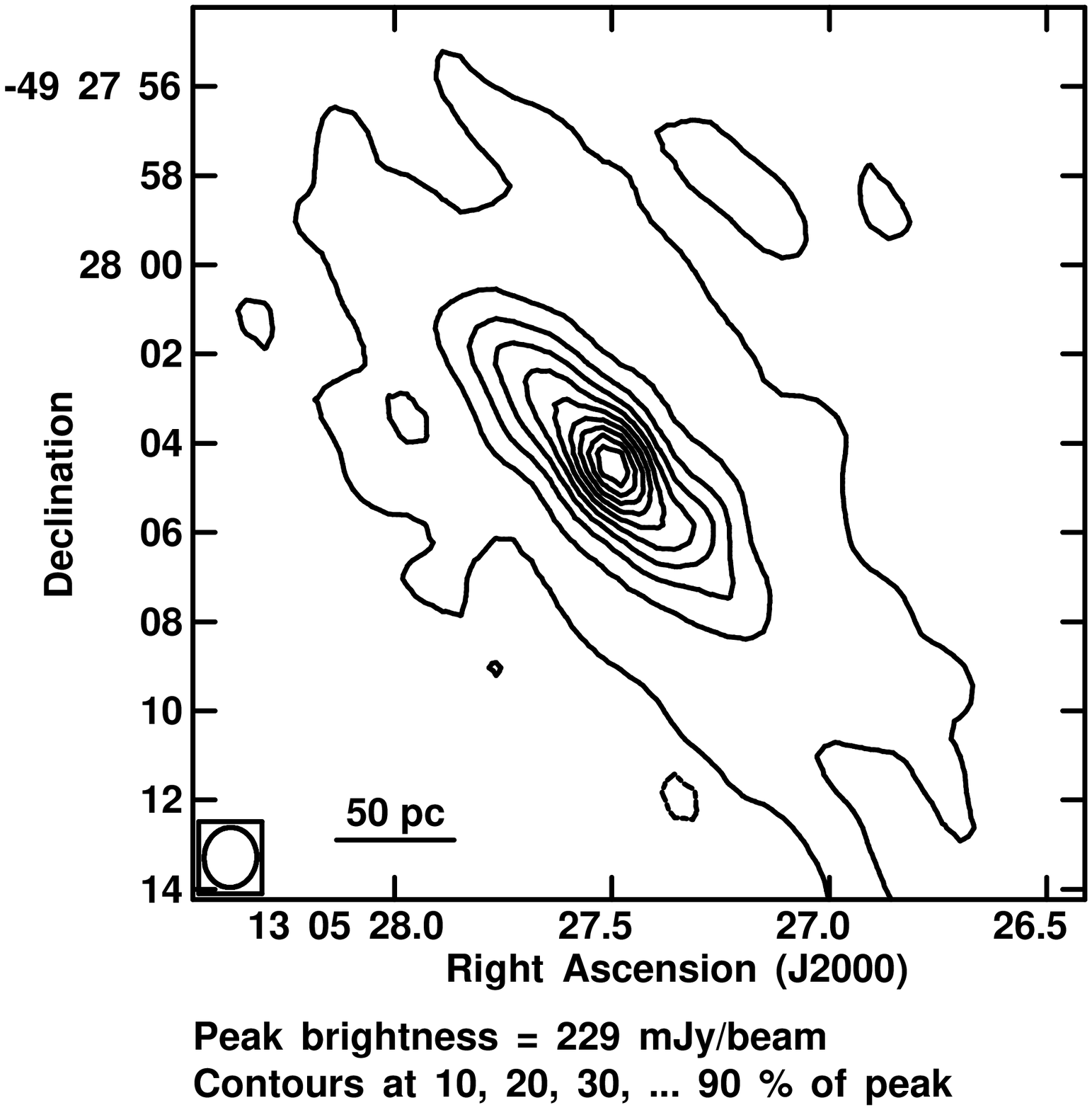}
\includegraphics[width=5cm]{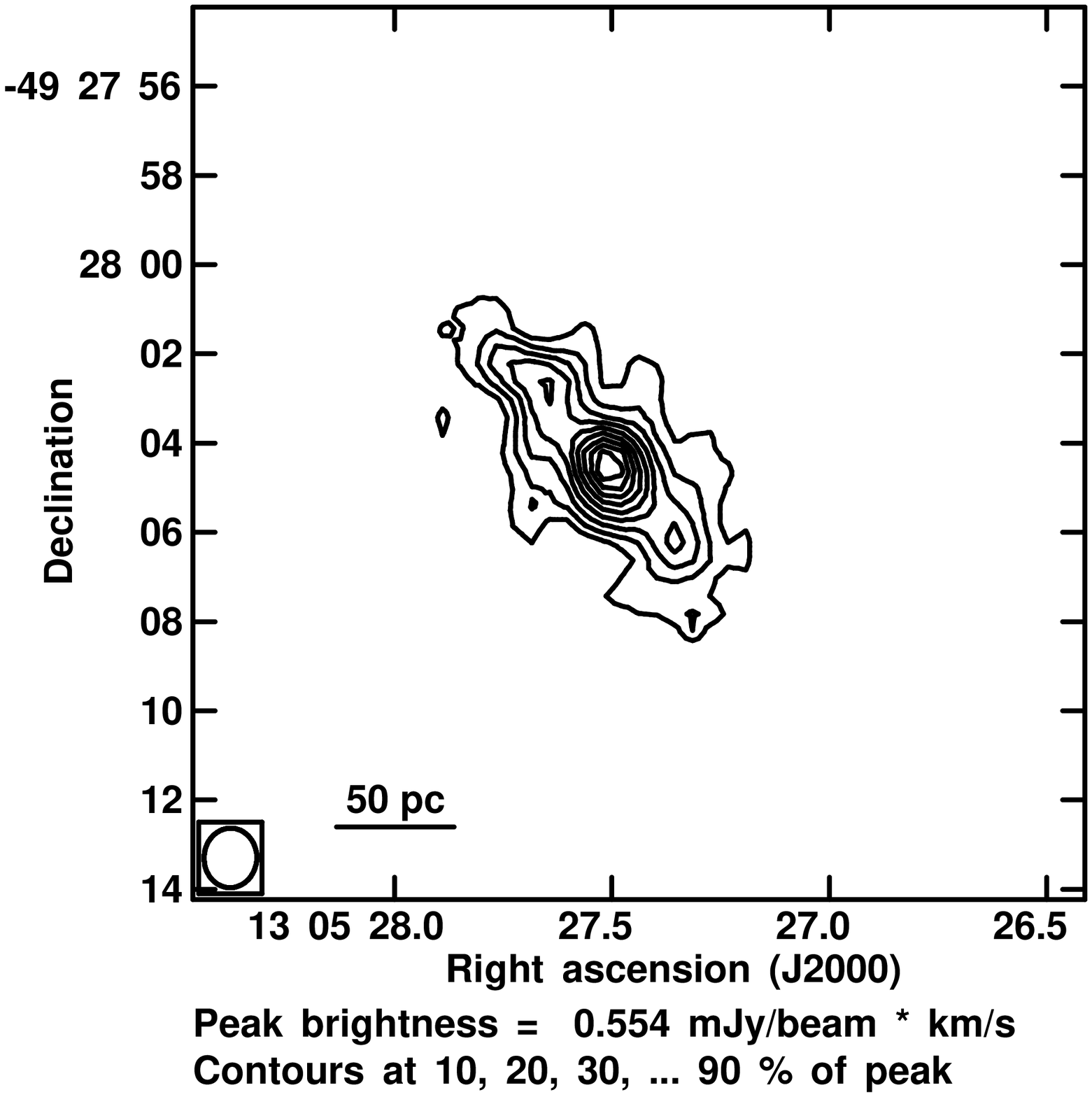}
\includegraphics[width=5cm]{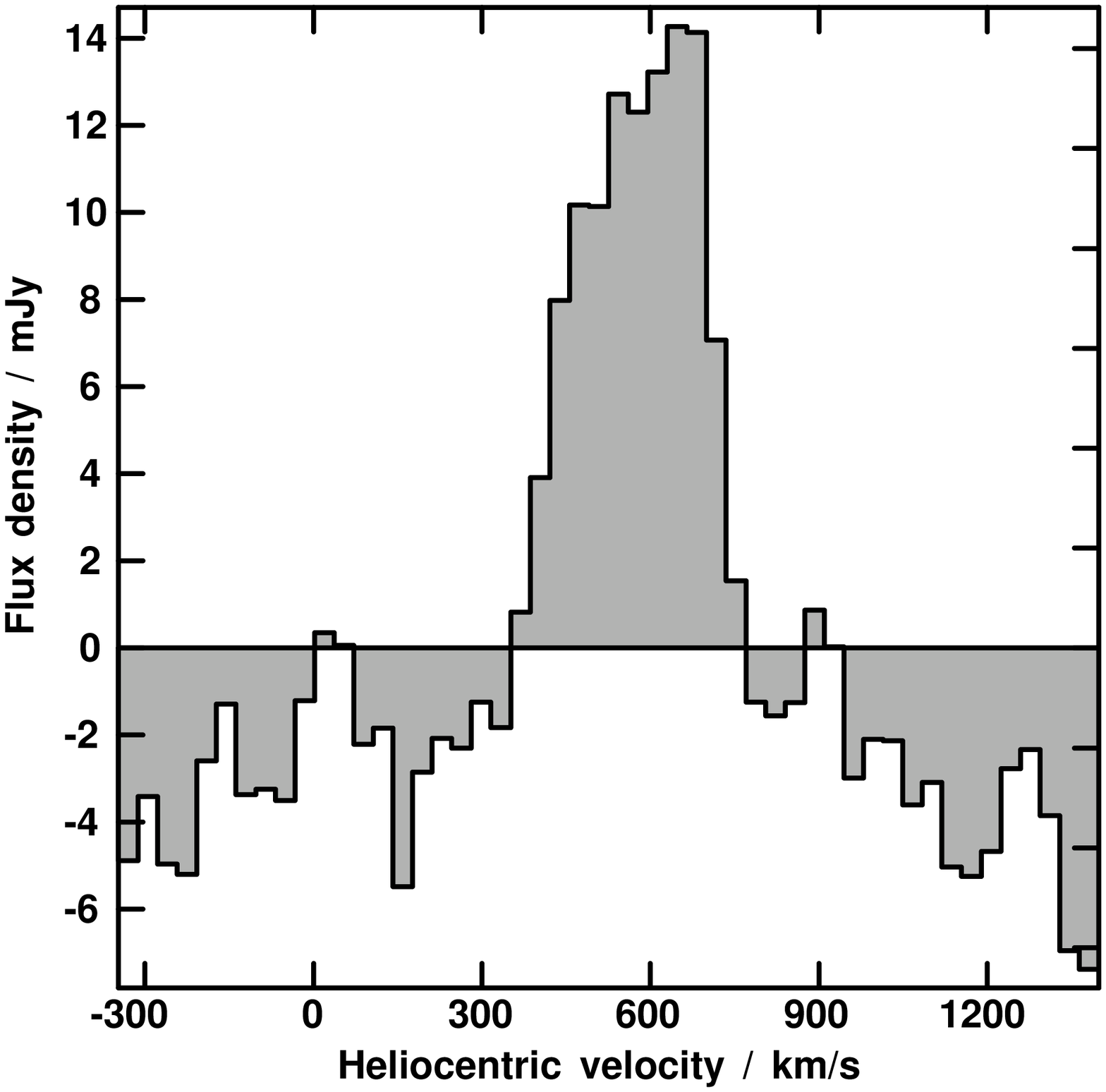}
\caption{
Left: ATCA 8.3 GHz + 8.6 GHz continuum image with uniform weight 
of NGC 4945 observed
for 22.7 h on 1993 Jul 25 + 1994 Sep 22 + 1994 Oct 04 (contours).  Middle: 
ATCA zeroth-moment image of H91$\alpha$ + H92$\alpha$ emission. 
Beamsize is $1.4'' \times 1.2''$.
Right: ATCA H91$\alpha$ + H92$\alpha$ line profile integrated over
line-emitting region in the zeroth-moment image. RMS noise is 
0.16\,mJy\,beam$^{-1}$\,channel$^{-1}$.
}
\end{figure}

\begin{figure}
\includegraphics[width=5cm]{roy_f3a.eps}
\includegraphics[width=5cm]{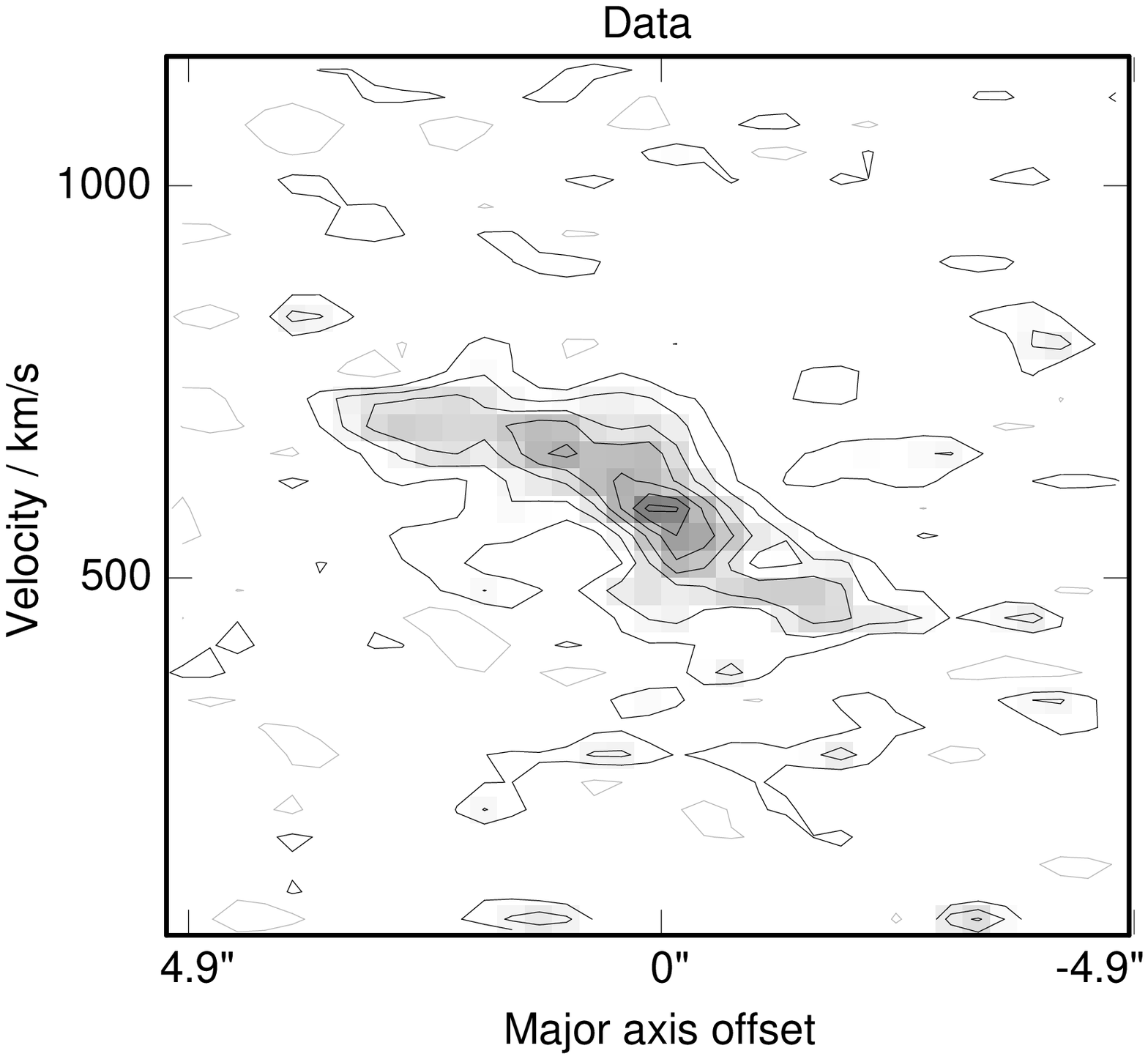}
\includegraphics[width=5cm]{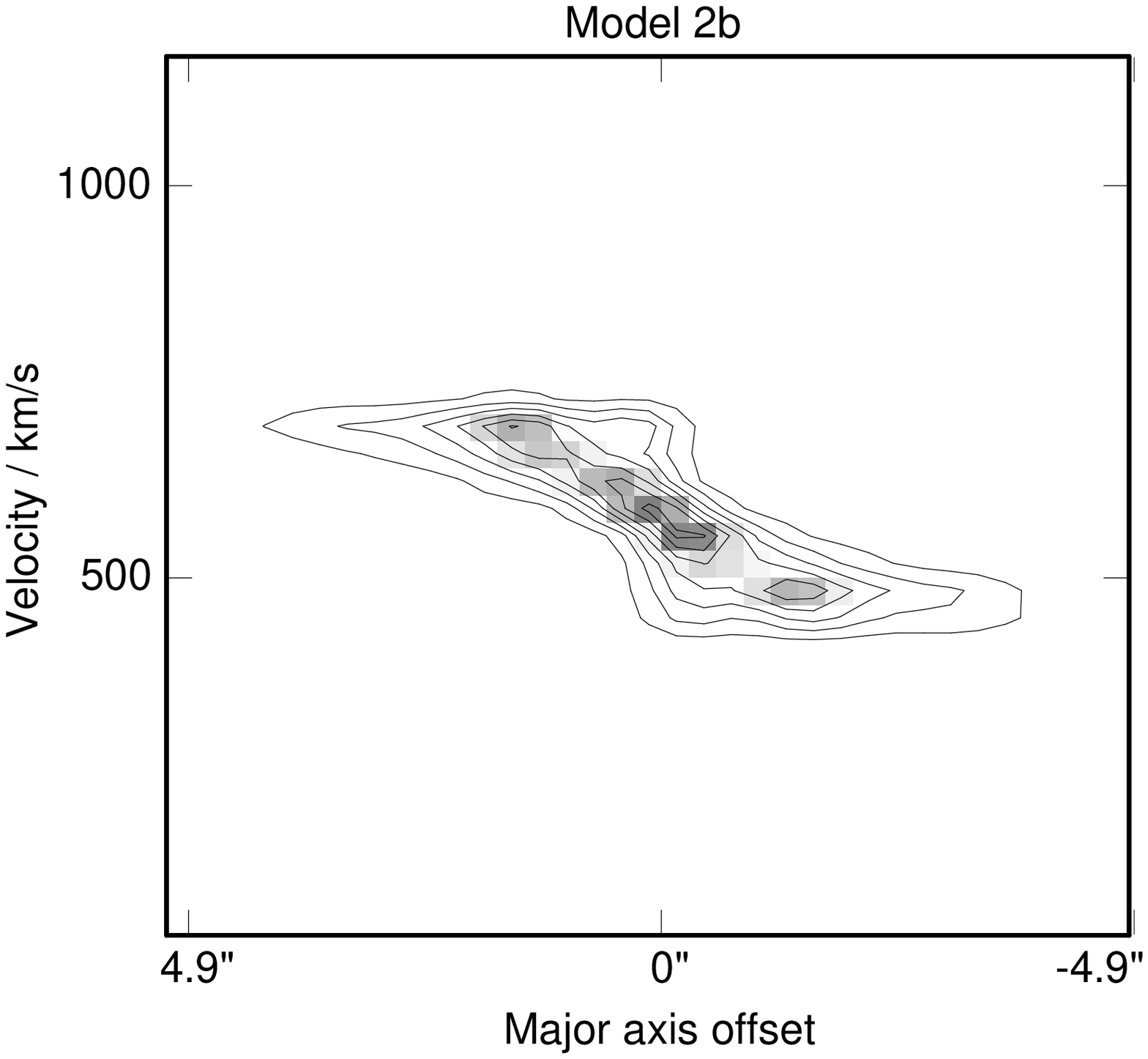}
\caption{
Left: ATCA first-moment image of H91$\alpha$ + H92$\alpha$ emission
showing rotation.  Beamsize is $1.4'' \times 1.2''$.
Middle: H91$\alpha$ + H92$\alpha$ rotation curve along the major axis
of the line-emitting region.
Right: Rotation curve from the best-fit dynamical model
}
\end{figure}

We used the integrated line strength (11.0~mJy), line width (280~km~s$^{-1}$),
size of the line-emitting region (70 pc), continuum emission (1439 mJy), and
spectral index (-0.75) to constrain conditions in the ionized gas.  Using the
collection of H~II regions model, models with 10 to 300 H~II regions, all with
$T_{\mathrm{e}} \sim 5000$~K, $n_{\mathrm{e}} \sim 10^{3}$~cm$^{-3}$ to
$10^{4}$~cm$^{-3}$ and size of (2 to 100)~pc produced good matches to
the line and continuum emission.

The inferred mass of ionized gas is $2\times10^{5}~M_{\odot}$ to
$6\times10^{5}~M_{\odot}$, depending on the model conditions, which requires a
Lyman continuum flux of $6\times10^{52}$~s$^{-1}$ to $3\times10^{53}$~s$^{-1}$
to maintain the ionization.  This flux is equivalent to the Lyman continuum
output of 2000 to 10000 stars of type O5, which infers a star-formation rate
of (2 to 8) $M_\odot$ yr$^{-1}$ when averaged over the lifetime of OB stars.

We fitted to the rotation curve a simple model consisting of a set of rings
which were coplanar and edge-on.  The brightness of each ring was determined
by deprojecting the observed zeroth-moment image to derive the radial distribution
of the line intensity.  The radial distribution showed a central peak and a
ring of emission 2.5'' (50~pc) from the nucleus.

We refined the model iteratively, varying the systemic velocity, rotation
velocity, and velocity dispersion until we achieved a reasonably close match
to the data, and found no reason to depart from a simple flat rotation curve.
We refined the radial profile of line emission strength,
increasing the central peak at the systemic velocity to 25 times the
brightness of the more extended emission.  The final model had a flat rotation
curve with $v_{\rm{systemic}} = 581$~km~s$^{-1}$, $v_{\rm{rotation}}$ =
120~km~s$^{-1}$, and $v_{\rm{dispersion}}$ = 15~km~s$^{-1}$.  We did not need
to invoke a bar or radial motion, though the data do not strongly exclude
either.

The rotation velocity of 120~km~s$^{-1}$ within the central 1'' (19~pc) radius
infers an enclosed mass of $3 \times 10^7 M_\odot$.  The water masers
(\cite{GreenhillEtAl1997}) infer an enclosed mass of $1 \times 10^6 M_\odot$
within 0.3~pc radius, and so most of the $3 \times 10^7 M_\odot$ is extended
between 0.3\,pc and 19\,pc radius of the nucleus.  The average surface density
within the central 19\,pc is $25000~M_\odot~$pc$^{-2}$, which 
exceeds the threshold gas surface density for star-formation of (3 to
10)~$M_\odot~$pc$^{-2}$ (\cite{Kennicutt1989}) by four orders of magnitude.

The 50~pc ring might be a Lindblad resonance, which would infer a bar.  The
bright peak in the line emission at the nucleus might be recombination-line
maser emission occuring along the line of sight towards the nucleus.  In that
direction, the gas velocity is transverse to the line of sight and so gas at
all distances share a common velocity, which maxizes the path length over
which RRL maser amplification can occur.

\section{Detection of the Circinus Galaxy}

The Circinus Galaxy, at a distance of 4.2 Mpc\cite{FreemanEtAl1977}, is 
situated behind the Galaxy
and evaded discovery until relatively recently \cite{FreemanEtAl1977}.  It
hosts a nuclear starburst and an obscured Seyfert nucleus, and is a water
megamaser source \cite{GardnerWhiteoak1982}.

Our ATCA observations discovered RRL emission, which is shown in Fig 4.

\begin{figure}
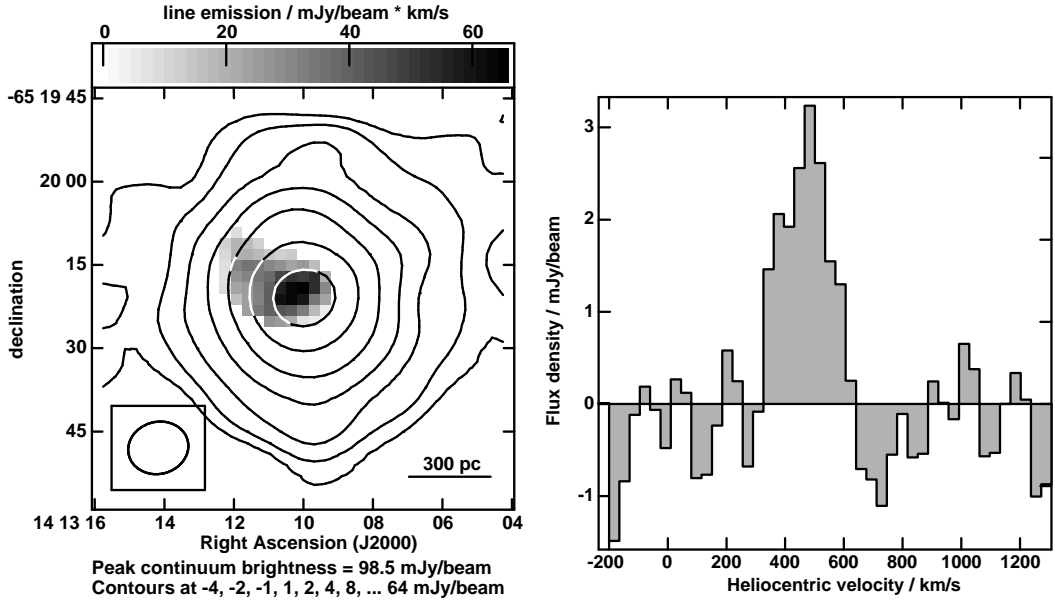

\includegraphics[width=7cm]{roy_f4a.eps}
\includegraphics[width=7cm]{roy_f4b.eps}
\caption{
Left: ATCA 8.4 GHz continuum image of the Circinus Galaxy observed for 37 h on
1993 Jul + 1994 Oct + 1994 Nov
(contours), superimposed on the grey scale zeroth-moment image showing
H91$\alpha$ + H92$\alpha$ line emission.  Beamsize is
$11.0'' \times 9.4''$, rms noise is 0.12\,mJy\,beam$^{-1}$\,channel$^{-1}$.
Right: ATCA H91$\alpha$ + H92$\alpha$ line profile integrated over the
line-emitting region in the Circinus Galaxy.
}
\end{figure}

We used the observed line
strength (3.2\,mJy), line width (280\,km\,s$^{-1}$), size of the 
line-emitting region (250\,pc), continuum emission (225\,mJy), and 
spectral index (-0.65) to constrain conditions in the ionized gas.  The
collection of H~II regions model with 15 to 10000 H~II regions, all 
with $T_{\mathrm{e}} \sim 5000$~K, $n_{\mathrm{e}} \sim (500$ to
$5 \times 10^{4})$~cm$^{-3}$ and size of (3 to 50)~pc produced good 
matches to the line and continuum emission.

The inferred mass of ionized gas is $3\times10^{3}~M_{\odot}$ to
$1\times10^{6}~M_{\odot}$, depending on the model conditions, which requires a
Lyman continuum flux of $1\times10^{52}$~s$^{-1}$ to $3\times10^{53}$~s$^{-1}$
to maintain the ionization.  This flux is equivalent to the Lyman continuum
output of 300 to 2000 stars of type O5, which infers a star-formation rate
of (0.2 to 2) $M_\odot$ yr$^{-1}$ when averaged over the lifetime of OB stars.

\section{Conclusion}

Seven galaxies were observed with the ATCA and one also with the VLA.  These
yielded a spectacular detection of NGC\,4945, and good detections of NGC\,3256
and the Circinus Galaxy.  Four other galaxies were not detected.  From the
detections we derived physical conditions and kinematics in the ionized gas in
the nuclear starbursts.

Future observations at high frequencies where RRLs are stronger and resolution
is higher will provide measurements of multiple transitions to provide tighter
constraints on the gas conditions, and the upcoming sensitivity improvements
coming with the EVLA, ALMA and SKA will make tremendous impact on this work.

%%%%%%%%%%%%%%%%%%%%%%%%%%%%%%%%%%%%%%%%%%%%%%%%
%% BACKMATTER
%%%%%%%%%%%%%%%%%%%%%%%%%%%%%%%%%%%%%%%%%%%%%%%%

%% If the bibliography is produced without BibTeX comment out the

\bibliographystyle{aipproc}   % if natbib is available

\bibliography{roy}

\IfFileExists{\jobname.bbl}{}
 {\typeout{}
  \typeout{******************************************}
  \typeout{** Please run "bibtex \jobname" to optain}
  \typeout{** the bibliography and then re-run LaTeX}
  \typeout{** twice to fix the references!}
  \typeout{******************************************}
  \typeout{}
 }

\end{document}